\documentclass[12pt,a4paper]{article}
\usepackage{amssymb}
\usepackage{amsmath}
\usepackage{cite}
\usepackage{graphics}
\usepackage[dvips]{color}
\usepackage[dvips]{graphicx}

\newcommand{\beq}{\begin{equation}}
\newcommand{\bea}{\begin{eqnarray}}
\newcommand{\bec}{\begin{center}}
\newcommand{\eeq}{\end{equation}}
\newcommand{\eea}{\end{eqnarray}}
\newcommand{\eec}{\end{center}}
\newcommand{\vev}[1]{\ensuremath{\langle #1 \rangle}}

\newcommand{\so}{\ensuremath{SO(10) \;}}
\newcommand{\pfrac}[2]{\genfrac{(}{)}{}{}{#1}{#2}}
\newcommand{\wur}{\dfrac{1}{2 \, \pi}}
\def\ga{\mathrel{\raise.3ex\hbox{$>$\kern-.75em\lower1ex\hbox{$\sim$}}}}
\def\la{\mathrel{\raise.3ex\hbox{$<$\kern-.75em\lower1ex\hbox{$\sim$}}}}
\newcommand{\mgut}{\ensuremath{M_{GUT}\,}}
\newcommand{\meff}{\ensuremath{M_{eff}\,}}
\newcommand{\meffth}{\ensuremath{M_{eff}(th)\,}}
\newcommand{\meffexp}{\ensuremath{M_{eff}(exp)\,}}

\newcommand{\astrong}{\ensuremath{\alpha_{strong}\,}}

\begin{document}

\begin{titlepage}  
\begin{flushright} 
\parbox{4.6cm}{UA-NPPS/BSM-10/02\\
              }
\end{flushright} 
\vspace*{8mm} 
\begin{center} 
{\large{\textbf {
Collective treatment of High Energy Thresholds \\
in SUSY - GUTs .
}}}\\
\vspace{14mm} 
{\bf A. ~\ Katsikatsou}  

\vspace*{6mm} 
  {\it University of Athens, Physics Department,  
Nuclear and Particle Physics Section,\\  
GR--15771  Athens, Greece}

\end{center} 
\vspace*{15mm} 

\begin{abstract}
Supersymmetric GUTs are the most natural extension of the Standard model unifying electroweak and strong forces. Despite their indubitable virtues, among these the gauge coupling unification and the quantization of the electric charge, one of their shortcomings is the large number of parameters used to describe the high energy thresholds (HET), which are hard to handle.
We present a new method according to which the effects of the HET, in any GUT model, can be described by   fewer parameters that are randomly produced from the original set of the parameters of the model. In this way, regions favoured by the experimental data are easier to locate, avoiding a detailed and time consuming exploration of the parameter space, which is multidimensional even in the most economic unifying schemes. To check the efficiency of this method, we directly apply it to a SUSY SO(10) GUT model in which the doublet-triplet splitting is realized through the Dimopoulos-Wilczek mechanism. We show that the demand of gauge coupling unification, in conjunction with precision data, locates regions of the parameter space in which values of the strong coupling \astrong are within the experimental limits, along with a suppressed  nucleon decay, mediated by a higgsino driven dimension five operators, yielding lifetimes that are comfortably above the current experimental bounds. These regions open up for values of the  SUSY breaking parameters $m_0, M_{1/2} \, <  \, 1 \; TeV \,$ being therefore accessible to LHC.
 
\end{abstract} 
\vspace{2.5cm}

\hspace{1cm}\hrule width 7.5 cm 

\vspace{.2cm}

\hspace{.2cm} {E-mail\,:\, kkatsik@phys.uoa.gr}

\end{titlepage} 
 \clearpage
\newpage

\baselineskip=17pt

\section{Introduction}\label{model}



Grand unified theories (GUTs) provide a simple and elegant framework for the unification of strong, weak and electromagnetic forces. In addition, they offer a simple explanation of the electric charge and hypercharge assignments to the quarks and leptons in the Standard Model and combine its seekingly unrelated  left and right - handed multiplets (five per family) into common representations of the larger unifying group. 
Moreover, their minimal supersymmetric versions, SUSY GUTs, lead to a successful gauge coupling unification 
at scales $\mgut \, \approx \, 10^{16}$ GeV, which is impossible to realize without supersymmetry. Also, in the framework of particulate SUSY-GUTs, the lightness of the neutrinos can be explained and a mechanism for Baryogenesis through thermal Leptogenesis is offered as an alternative to Baryogenesis through electroweak phase transition, which requires large CP - violating phases. 

In this note, our goal is to check the viability of SUSY GUTs, using electroweak precision and proton decay data by developing an integrated and simple scheme for the treatment of high energy thresholds (HET), which can be applicable, in principle, to any GUT model. In this scheme, the HETs are collectively parametrized by a small number of  properly chosen variables, which are therefore easier to handle in phenomenological analyses. These are produced from the numerous parameters defining the HETs in a random way. 

We employ our method to a SUSY - \so  \cite{Georgi},  
which seems to be a very promising candidate for a unified description. Models based on 
\so  unify all quarks and leptons of one family into one irreducible spinor representation, 
they give upper bounds on proton lifetime that still survive the current experimental bounds,
they naturally incorporate the see-saw mechanism, reproducing the current neutrino oscillation data and explaining the lightness of the left-handed neutrinos with the sterile neutrino mass being of order \mgut \cite{neutrinos}, 
and finally predict Yukawa unification \cite{Yukawauni}, $\lambda_t \, = \, \lambda_b \, = \, \lambda_\tau  \, = \, \lambda_{\bar{\nu} _\tau} $, for large $\tan \beta $.
Our analysis is based on a minimal supersymmetric \so  model first proposed in \cite{Barr:1}, whose low energy effective theory is the constrained MSSM. 

As we have already noted, a common characteristic of all \so models is the inclusion of all quarks and lepton of one generation, along with the right handed neutrino, in the same spinorial, $\mathbf{16}$, representation. 
The \so generators belong to the adjoint representation $\mathbf{45}_V$.  

As far as the content of the Higgs sector is concerned, there are two approaches that have been considered in the literature. Their difference rely on the way the Higgs mechanism is realized at the GUT scale. The first approach uses pairs of spinor Higgs multiplets in $\mathbf{16}_H \, + \, \overline{\mathbf{16}_H}$ representations to reduce the rank of the group \cite{Barr:1, Babu, Barr:2}, whereas the other adopts a pair of $\mathbf{126}_H \, + \, \overline{\mathbf{126}_H}$ instead \cite{Mohapatra}. In both cases, a Higgs field in the $\mathbf{45}$ or the $\mathbf{54}$ representation is also needed, in addition, to further break \so \, to the Standard Model. Moreover, in order to build a viable model in \so, one should always take into account that the Higgs multiplets of \so should include the two MSSM Higgs doublets $H_u, \; H_d$, which give masses to the up and down quarks respectively and additional Higgs fields, which are necessary to obtain the appropriate symmetry breaking pattern at the unification scale. The Higgs content of the particular model, which we will study, follows the first approach and is characterized as minimal. It consists of one Higgs multiplet $A$ in the adjoint $\mathbf{45}_H$ representation, two pairs of $\mathbf{16}_H \, + \, \overline{\mathbf{16}}_H$ multiplets, named $C \, + \, \overline{C}, \; C' \, + \, \overline{C'}$,  and two Higgs $T_1, \, T_2$ in the vector $\mathbf{10}_H$ representation.

\begin{figure}[h]
\begin{center}
\rotatebox{0}{\includegraphics[]{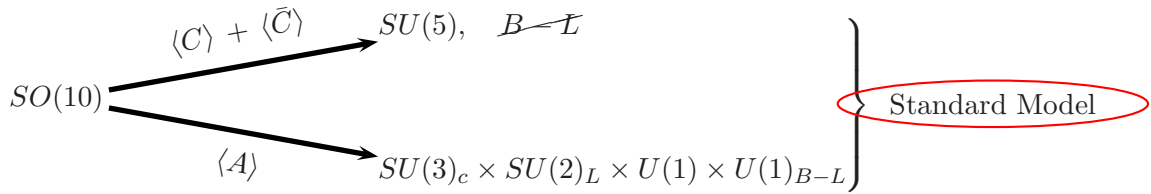}} 
\caption[]{\small The breaking scheme of \so to Standard Model.}
\label{breaking}
 \end{center}
\end{figure}

\so \, is spontaneously broken to the Standard Model without intermediate breaking scales, as shown in 
Figure \ref{breaking}. 

\vspace*{5mm}
%
%
%
%
In general, there are at least four sectors that are needed in the Higgs superpotential to accomplish the \so breaking \cite{Barr:1, Barr:2}: the doublet-triplet-splitting sector $W_{2/3}$, the adjoint sector $W_A $, the spinor sector $ W_C $ and the adjoint-spinor coupling sector $W_{ACC'} $, so that the total superpotential is 
$$W \, = \, W_A \, + \, W_C \, + \, W_{ACC'} \, + \, W_{2/3}  \quad .$$
The issue of the doublet-triplet splitting in \so is, in fact, a manifestation of the gauge hierarchy problem which is present in every grand unified theory. In SUSY $SO(10)$, the two Higgs doublets, $H_u, \; H_d$, required for the electroweak symmetry breaking, are contained in the vector Higgs multiplets, $H_u \, \subset \, \mathbf{5}, \; H_d \, \subset \, \mathbf{\overline{5}}$ of $SU(5)$, which fall into the 
$ \; \mathbf{10}_H \;   $ representation of $SO(10)$.

This pair of weak doublets must remain massless after the \so breaking at the GUT scale, while their color-triplet partners in the vector Higgs multiplet should obtain superheavy masses. The doublet-triplet splitting is implemented by the Dimopoulos - Wilczek mechanism \cite{DWmechanism}. This mechanism assumes the existence of a term of the form $ T_1 \, A \, T_2$ in the Higgs superpotential and demands the adjoint Higgs multiplet $A$ to have a vev along the $B-L$ direction,
\beq
\vev{A} \, = \, diag(\alpha, \,\alpha, \,\alpha, \, 0, \, 0) \, \otimes \, i \tau_2. 
\label{vevA}
\eeq
The parameter $\alpha$ is of the order of the GUT breaking scale $M_{GUT}.$ If one adds in the superpotential the mass term $M_2 \, (T_2)^2$, half of the four Higgs doublets contained in $ T_1, \, T_2$ remain massless and thus the picture of the MSSM Higgs spectrum is revealed at the low energy  effective theory. Hence, the doublet-triplet-splitting sector will be assumed to have the following form, 
\beq
W_{2/3} \, = \, \lambda T_1 \, A \, T_2 \, + \, M_2 \, T^2_2,
\eeq
where $\lambda$ and $M_2$ are the massless and GUT scale massive parameters, of this sector, respectively.

The adjoint sector, $W_A$, is responsible for the Dimopoulos - Wilczek vev $ \vev{A}$ given in (\ref{vevA}), which breaks $SU(5)$ to the Standard Model symmetry preserving $U(1)_{B - L}$ ( see Figure \ref{breaking} ).  

The spinor sector, $W_C$, of the Higgs superpotential forces the pair of spinor Higgs multiplets $C \, + \, \overline{C}$ to get superheavy vevs along the $SU(5)$ singlet direction. In this way, the rank of the group is reduced from 5 to 4, since the $B - L$ symmetry is broken. 

Finally, the adjoint-spinor coupling terms, $W_{ACC'}$, are necessary to prevent the manifestation of colored pseudo-Goldstone bosons with small masses, which may destroy the unification of gauge coupling constants and the low energy particle spectrum.

Apart from providing a decent Higgs mechanism, the model just described yields reliable results for the fermion masses \cite{Barr:2, Pati}, including neutrino masses and their oscillations \cite{Barr:2, Pati}, at the electroweak scale by introducing the proper Yukawa terms at the GUT scale. 

This paper is organized as follows.
In Sec.\ref{secproton}, we briefly discuss the proton instability. The expected proton lifetime depends on a superheavy mass \meff which can be expressed in an efficient way as a function of HETs and the  gauge coupling constants at $M_Z$.
In Sect.\ref{sechet}, we propose a new method to count in the effects of the HETs in the running of the gauge couplings, needed for finding \meff,  imposing the experimental constraints from \astrong and proton decay.
The results of our analysis are presented in Sec.\ref{results1} and our conclusions are given in ~{Sec.\ref{conclusions}.}

\section{Proton Decay}
\label{secproton}

In every GUT model, the baryon number violation is unavoidable and consequently nucleons decay. In \so \, 
proton decays through  D=5 operators, which are induced via the exchange of superheavy color triplet Higgsinos in the $\mathbf{10}$ Higgs multiplet \cite{Dimopoulos, Ellis, Nath, Yanagida}, with the dominant decay mode being $p \, \rightarrow \, \overline{\nu} \, K^+$. These operators arise from an effective superpotential which is inverse proportional to a mass parameter to be denoted by
\beq
M_{eff} \, = \, \frac{M_3 \, M_3'}{M_2} \, = \, \frac{(\lambda \alpha)^2}{M_2} \, ,
\label{meff}
\eeq
where $M_3 \, M_3'$ are the superheavy color triplet Higgs masses of the vector Higgs multiplets $T_1, \, T_2$. 
Thus, the expected proton lifetime turns out to be proportional to $M_{eff}^2$. 

For the dominant decay mode, the decay rate is:
\beq
\Gamma( p \, \rightarrow \, \bar\nu \, K^+ ) \; = \; \sum_{i \, = \, e, \, \mu, \, \tau} \Gamma( p \, \rightarrow \, \bar\nu_i \, K^+ ) \, .
\label{pdwidth} 
\eeq
Each of the partial widths in (\ref{pdwidth}) are given by \cite{Nath1, Nath, Yanagida}:
\beq
\Gamma( p \, \rightarrow \, \bar\nu_i \, K^+ ) \; = \; \biggl( \dfrac{\beta_p}{M_{eff}} \biggr)^2  \; |A|^2 \; |B_i|^2 \; C \, .
\label{width}
\eeq
The factor $\beta_p$ denotes the hadronic matrix element between the proton and the vacuum state of the 3 quark operator \cite{bp1}, employed in the appropriate chiral Lagrangian schemes \cite{Claudson}.
In our approach, we use \cite{gavela} 
$$\beta_p \, = \, (5.6\pm 0.5) \, \times \, 10^{-3} \, \mathrm{GeV}^3 \,  $$
extracted from lattice gauge calculations. $A$ in (\ref{width}) depends on quark masses (at 1 GeV) \cite{Ellis} and CKM matrix elements and is given by
\beq
A \, = \, \frac{\alpha_2^2}{2 M_W^2} \, m_s \, m_c \, V_{21}^{\dagger} \, V_{21} \, A_S \, A_L.
\eeq
The first parameter, $A_S$, represents the short-range renormalization effects between GUT and SUSY breaking scales \cite{Ellis, Yanagida}, while the second one, $A_L$, accounts for 
the long-range renormalization effects between SUSY scale and 1 GeV \cite{Ellis}. Their values are
$A_S \, \simeq  \,0.93$, for $m_t \, = \, 175$, \cite{As} and $A_L \, \simeq  \, 0.32 \;$ (2-loop result),   
\cite{Pati}.

In (\ref{width}), the $B_i$s are the functions that describe the dressing of the loop diagrams and are given by the formula
\beq
B_i \, = \, \frac{1}{\sin 2\beta} \; \frac{m_i^d \, V_{i1}^{\dagger}}{m_s \, V_{21}^{\dagger}} \; 
\biggl( P_2 \, B_{2i} \, + \, P_3 \, \frac{m_t \, V_{31} \, V_{32}}{m_c \, V_{21} \, V_{22}} \, B_{3i} \biggr)
\, ,
\label{bi_prtn}
\eeq
%
where $B_{ji}$ is the contribution of the $j$th generation particles in the loop with
\beq
B_{ji} \, = \, F(\tilde u_i, \tilde d_j, \tilde W) \, + \, (\tilde d_j \rightarrow \tilde e_j) \, .
\label{bjila}
\eeq
The functions $F$ in (\ref{bjila}) contain the corresponding loop integrals \cite{Nath1}, with $i=1, 2, 3$ and $j=2, 3$, while  $P_2$ and $P_3$ are inter-generational, CP violating phases given by \cite{Nath1}. 
\beq
P_i \, = \, e^{i \, \gamma_i}, \quad \sum_i \gamma_i \, = \, 0 \; , \quad i=1,2,3 \, .
\eeq
%

There exist two distinct limiting cases having to do with the relative contributions of the second and third generation: the destructive interference, occurring for $P_3/P_2 \, = \, - \,1$, and the constructive one when $P_3/P_2 \, = \, + \,1$. We adopt the second case to achieve maximum mixing, and hence smaller, lifetimes that are more tightly constrained by data. 

Finally, the factor $C$ in (\ref{width}) contains chiral Lagrangian factors, which convert a Lagrangian involving quark fields to the effective Lagrangian involving mesons and baryons \cite{C}.
Its value has been calculated to be 
$C \, = 1.014 \,$, according to the values given in \cite{Nath1}.

The current experimental lower bound on proton lifetime from Super-Kamiokande is \cite{pdb}
\beq
\tau_{(p \, \rightarrow \, \overline{\nu} \, K^+)} \, > \, 1.6 \times 10^{\, 33} \, \mathrm{yrs} \, .
\label{plifetime}
\eeq
For given SUSY inputs, this constrains the value of the parameter $M_{eff}$, which, as we shall see, 
depends on the values of the gauge couplings at $M_Z \,$ and other high energy threshold parameters of the theory.
Note that in ref.\cite{Yanagida}, in the context of the $SU(5)$ model, the color-triplet Higgs boson mass, which is the analogue of $M_{eff}$ in $SU(5)$, is constrained by the precision measurement bounds put on \astrong. 
Following an analogous treatment as in \cite{Yanagida}, we first solve the 1-loop RGEs for the gauge coupling constants ${\alpha}_i$, in the $\overline{DR}$ scheme, from the GUT scale $M_{GUT}$ down to the electroweak scale $M_Z$ and express \meff in terms of ${\alpha}_i (M_Z)$. In this approach, we count in the high energy thresholds \cite{Weinberg,Hall} of the superheavy spectrum, as well as the low energy thresholds of all sparticles and heavy SM particles. The resulting expression, for the \so \, is
\beq
\frac{M_{eff}}{M_Z} \, = \,\, \mathrm{e}^{h ( {\alpha}_i^{-1} )} \,\, f(x) \, ,
\label{yan_ours}
\eeq
with
$$h ( {\alpha}_i^{-1} ) \, = \, \dfrac{5 \, \pi}{6} \, \bigl[ \, 3 \, {\alpha}_2^{-1} (M_Z) \, - \, {\alpha}_1^{-1} (M_Z) \, - \, 2 \, {\alpha}_3^{-1} (M_Z) \, \bigr] \, .$$
The effect of the low energy thresholds in (\ref{yan_ours}) are encoded within $\alpha_i$ from the low energy boundary conditions that are imposed at $M_Z$. In fact 
\bea
{\alpha}_1^{-1}(M_Z) \,&=&\, \dfrac{3}{5} \, \alpha_{em}^{-1} \, \cos^2 \theta_w \,
(\,1 - \Delta_\gamma + \dfrac{ \alpha_{em} }{ 2 \, \pi } \, \ln \dfrac{M_S}{M_Z} \,) 
\nonumber \\
{\alpha}_2^{-1}(M_Z) \,&=& \, \alpha_{em}^{-1} \, \sin^2 \theta_w \,
(\,1 - \Delta_\gamma + \dfrac{ \alpha_{em} }{ 2 \, \pi } \, \ln \dfrac{M_S}{M_Z} \,) 
\nonumber \\
{\alpha}_3^{-1}(M_Z) \,&=& \, \alpha_{strong}^{-1}(M_Z) \,|_{\overline{MS}} \, - \, \dfrac{1}{4 \, \pi}
+\, \dfrac{1}{2 \, \pi} \,\ln \dfrac{\tilde{M}_S}{M_Z} 
\label{allit}
\eea
where $\, \alpha_{em} \, , \; \alpha_{strong} \,$ are the electromagnetic and strong coupling constants respectively, 
$\, \theta_w \,$ is the weak mixing angle and 
$\, {\tilde{M}}_S, \, M_S \,$ account for the low energy threshold corrections. For further details and for  the definition of the remaining quantities in (\ref{allit}) see \cite{Lahanas}. 
At a subsequent stage, one can include a correction factor to (\ref{yan_ours}) to take care of the small two-loop corrections to the gauge coupling running. 
In our numerical analysis, the two-loop effects are properly counted for since we integrate the 2-loop RGEs of the gauge coupling constants in the $\overline{DR}$ scheme. 
More details on the method that we employ will be presented in the following section. 

The function $f(x)$ in (\ref{yan_ours}) depends only on GUT physics details and, in particular, on the high energy thresholds of the superheavy particles involved. For the case of \so , it is found that, 
\beq
f (x) \, = \, \dfrac{9}{16 \, \sqrt{2}} \biggl[ \pfrac{1 \, + \, 8 \, x^2}{1 \, + x^2}^4 \; \dfrac{(1 \, + 4 \, x^2)^3}{1 \, + \, 32 \, x^2}\biggr]^{1/2}. 
\eeq
In this expression, the effect of the high energy thresholds of this particular model depend only on the massless, free parameter $x$, which is defined as
\beq
x \, \equiv \, \dfrac{\alpha}{ 2 \, c},
\label{sh_param}
\eeq
with $\alpha, \, c$ being connected to the GUT scale vevs of the adjoint and spinor Higgs fields respectively.
Obviously, the larger the $x$, the larger the function $f(x)$ is, facilitating the satisfaction of the proton decay bounds. However, $x$ is naturally expected to be of $\mathcal{O}(1) \,$  as being the ratio of vevs which are both of order  $\sim \mgut$. On these grounds, $x$ cannot be taken  arbitrarily large.

For comparison, in the $SU(5)$ model the function  $f (x) \, = \, 1$ and we recover exactly the result presented in \cite{Yanagida}.

\section{High Energy Thresholds} \label{sechet}

It is clear from eq.\ref{yan_ours} that, in order to derive \meff, we need the values of gauge couplings  at the electroweak scale. Their values are set by (\ref{allit}) in terms of  
$\, {\tilde{M}}_S, \, M_S \,$. These are not physical masses but rather a convenient device  which encodes all information for the low energy supersymmetric thresholds and heavy standard model states. Given the SUSY inputs and $\,\alpha_{em},\, \alpha_s, \, \sin^2 \theta_w $, we solve the RGEs 
imposing the gauge coupling unification condition,  
\beq
\alpha_1 (\mgut) \, = \, \alpha_2 (\mgut) \, = \, \alpha_3 (\mgut) \, \equiv \, \alpha_G \, ,
\label{unif}
\eeq
but we don't restrict further our analysis by insisting on Yukawa unification.
At \mgut, we also impose universal boundary conditions induced by gravity mediated SUSY-breaking for the soft supersymmetry breaking parameters, although other schemes may be available, 
$$
m_i \, = \, m_0 \quad , \quad  M_i \, = \, M_{1/2} \quad , \quad A_i \, = \, A_0 \,.
$$


Using standard procedure, we calculate the gauge coupling constants at any scale $\mu$ below the 
SUSY thresholds by solving the appropriate RGEs incorporating the effects of the thresholds of all heavy Standard Model particles of mass $m_{SM_i} > \mu$ as well as those of all low energy SUSY particles $S_i$ and 
superheavy particles $H_i$ with masses $\, \sim M_{GUT} \,$ associated with the specific GUT model at hand. The result is 
\bea
\alpha_i^{-1} (\mu) &=& \alpha_G^{-1} (\mgut) \, + \, (\mathrm{2-loops \, effects}) \nonumber\\
&+& \wur \,  (b_i^{SM} \, + \, b_i^{SUSY}) \, \ln \dfrac{\mgut}{\mu}  
\, + \, \wur \, \sum_{SM_i} \, b_i^{SM_i} \, \ln \dfrac{\mu}{m_{SM_i}}\label{genRGE}
\\
&+&
\wur \, \sum_{S_i} \, b_i^{S_i} \, \ln \dfrac{\mu}{m_{S_i}} \, + \, 
\wur \, \sum_{H_i} \, b_i^{H_i} \, \ln \dfrac{\mgut}{m_{H_i}} .\nonumber
\eea
$\, b_i^{A} \,$ are the beta function coefficients of any species $\,A \,$ with $\,b_i^{SM}, \; b_i^{SUSY}  \,$ 
being the beta function coefficients of $\, \alpha_i \,$ of the SM and low energy supersymmetric modes respectively. 

From the evolution of $\alpha_i$ from \mgut to the lowest high energy threshold,  $M_L$, 
one can easily derive, ignoring momentarily the two loop effects, that 
\beq
\alpha_i^{-1} (M_L) \, = \, \alpha_G^{-1} (\mgut) \, + \, 
\wur \, b^{GUT} \, \ln \dfrac{\mgut}{M_L} \, + \,
\wur \, \sum_{H_i} \, b_i^{H_i} \, \ln \dfrac{M_L}{m_{H_i}} \,  ,
\label{runmxml}
\eeq
where $b^{GUT} \, \equiv \, b^{SM } + b^{SUSY}+ b^H \,$. 
By defining
\beq
\alpha_G^{-1} (M_L) \, \equiv  \, \alpha_G^{-1} (\mgut) \, + \, 
\wur \, b^{GUT} \, \ln \dfrac{\mgut}{M_L} \, , 
\label{agutml}
\eeq
which represents the running from \mgut to $M_L$, if HETs are ignored, and by using 
\beq
c_i \, \equiv \, 
\wur \, \sum_{H_i} \, b_i^{H_i} \, \ln \dfrac{M_L}{m_{H_i}} 
\label{cidef}
\eeq
equation (\ref{runmxml}) can be cast in the form, 
\beq
\alpha_i^{-1} (M_L) \, = \, \alpha_G^{-1} (M_L) \, + \, c_i \, .
\label{MLcon}
\eeq
where the effect of HET is included within $c_i$.

Eqs. (\ref{MLcon}) can serve as boundary condition at the lowest HET, $M_L$ that takes into account the effects of all HETs, which are included within the  constant $\, c_i\,$.
Between $M_L$ and the electroweak scale $\mu \, = \, M_Z$ no high energy thresholds are present and RGEs run as usual.

The importance of the adoption of the parametrization (\ref{cidef})
lies on the fact that the three $c_i$'s carry all information on the the masses of the non-singlet superheavy fields of the model which contribute to the HETs in the running from $M_Z$ to $M_L$ and vice versa.
For any set of the model parameters, say $p_j$, we assign a \textquotedblleft vector\textquotedblright \, in a five dimensional space 
$\, \vec{c} = ( c_1, \, c_2, \, c_3,  \,  \, M_L,  \, M_{GUT} \, ) $, which includes, except $c_i$, the values of the maximum, \mgut, and lowest high energy mass, $M_L \,$. 
For instance, in the version of the \so model we are considering, the number of the parameters $p_j$ is ten and for any point in this ten-dimensional parameter space there correspond  twenty-five correlated superheavy masses from  which we determine $\, M_{GUT}, \, M_L   \,$ and $c_i \,$s through their definitions in (\ref{cidef}).
Further, in order to utilize $c_i$ as inputs, we use a random sample generator, which assigns random numbers to the GUT parameters $p_j$. In this way, random points $\vec{p} \equiv (p_1, p_2, ...p_N)  $ are drawn in the model parameter space and each of this is mapped to a $\, \vec{c} \,$ defined before. 
Consequently, our analysis is fully constrained from the random sample results and instead of dealing with a large number of GUT parameters and masses, we only have a few to consider in our analysis, namely 
$\, c_1, \, c_2, \, c_3,  \,  \, M_L,  \, M_{GUT} \, $,
which define  $\, \vec{c} \,$. 
The random  procedure  actually makes a selection by mapping the parameter space to a rather confined  region, at least in the \so model, which is 
spanned by the vectors $\, \vec{c} \,$. Then, within this region, points satisfying the experimental criteria  can be sought and the region shrinks even more. Consequently, one avoids time-consuming scans over a multidimensional (10-dimensional in \so ) parameter space, since the random  procedure  has already selected  the points $\, \vec{c} \,$ which meet the criteria. 
In this way we have found a very convenient way to parametrize the effect of HET, which is applicable to almost any GUT model, using the variables $\, \vec{c} \,$.

We will now discuss in more detail the numerical procedure we follow. 
For any input point in the space of the randomly generated vectors, $\, \vec{c}_{\, in} \,$, we pick up the points $c_1^{\, in}, \, c_2^{\, in} \,$ and   
by running the 2-loop RGEs upwards, starting from from $M_Z$, we 
determine the values of $M_L$ and $\alpha_G(M_L)$
where the boundary condition (\ref{MLcon}) are satisfied. 
By (\ref{agutml}) the value of $ \alpha_G(\mgut) $ is also determined. Note that in this way the input value, $M_L^{\, in} $, is not the same with the extracted value, $M_L\,$. Subsequently, we define 
$c_3\,$ so that  
\beq
c_3 \;=\;    \alpha_3^{-1} (M_L) \, - \, \alpha_G^{-1} (M_L)  \, .
\label{c3}
\eeq
and relation (\ref{MLcon}) is satisfied for the coupling $\alpha_3 \,$. That done, the 
initial point  $\, \vec{c}_{\, in} \,$, with $c_i^{\, in}, \, M_L^{\, in} $,   has  moved to 
another with coordinates  
$\, c_1^{\, in}, \, c_2^{\, in}, \, c_3,  \,  \, M_L,  \, M_{GUT}^{\, in} \,$, i.e. the values of 
$c_3\,$ and $M_L\,$ have been only changed.  
This procedure is repeated in each iterative step from the electroweak to the GUT scale, in the usual manner,  until convergence has been obtained. 
In each iteration step, $c_1, \, c_2 \,$ are also corrected to meet the unification criteria but these corrections are small. 
Thus, at the end, we get a 
final point, $\, \vec{c}_{\, fin} \,$, which is a successful point if it belongs to the set of the randomly generated vectors $\, \vec{c}\,$ or unsuccessful and hence discarded if it lies outside the region spanned by $\, \vec{c}\,$ points. 
In order to test the correctness of our method, we check, at the end, if a successful point maps to itself,  modulo small differences due to numerical accuracies. This procedure will be explained and quantified in more detail later. 

Obviously, the set of successful points that pass the unification test will be reduced if additional physics constraints are imposed, like, for instance, bounds put by proton decay.

The advantage of the method is that the running, from $M_Z $ to $M_L$, of the couplings involved is done without the effect of the HET in the RGEs. Their effect is taken into account by the boundary condition (\ref{MLcon}), which incorporates all the HET information within $c_i$'s.

The boundary conditions for the couplings and soft mass parameters are imposed at the unification scale, and, in the running from  $M_{GUT}$ to   $M_L$,  the HET play an essential role. Since, in our approach, we want to treat HET collectively, through quantities similar to $c_i \,$, without knowledge of their precise values, we use the two-loop RGEs without the contribution of the superheavy particles, which are known. However, in order to include the effect of the HETs,  at the end we correct each derived quantity at the scale $M_L$ as 
$$
F^{cor}(M_L) \,=\, F (M_L) \,+\, \Delta_F \, .
$$
The added  quantity, $\Delta_F \,$, depends on $c_i \,$, on the beta function coefficients 
$b_i^H \equiv \sum_{H_i} \, b_i^{H_i} $ of the superheavy modes that are known and on 
$\ln \,( M_{GUT} / M_L ) \,$. The correction $\Delta_F \,$ is different for each quantity F and, for its knowledge, the one-loop explicit dependence of the RGE of the $\, F \,$ need to be known. In particular, if 
$$
\dfrac{d \, F}{d \, ln Q} \,=\, \sum_i \, G_i \, \alpha_i + ... \quad ,
$$ 
where the ellipses denote contributions not explicitly dependent on the gauge couplings, the correction is 
$$
\Delta_F \,=\,  \dfrac{\alpha_G^2}{4\, \pi} \,  \ln \left( \dfrac{M_{GUT} }{ M_L} \right) \,\sum_i \,G_i(M_{GUT}) \,
\left[ \ln \, \left( \dfrac{M_{GUT} }{ M_L} \right) \, 
b_i^H \, + \, 2 \, \pi \, c_i \right] \, .
$$
This is a valid approximation provided that the lowest, $M_L$, and the highest, $M_{GUT}$, threshold are not far apart. In particular, $ \frac{M_L}{\mgut} \, < 10^{-3}$ is demanded. From the random samples, we find that  
on an average $\log \frac{M_L}{\mgut} \, \simeq  2.7$ and thus the approximation is more than satisfactory.

With these in mind, we calculate couplings and masses at the electroweak scale in the ordinary manner with the
values of the gauge couplings determined by the electroweak precision measurements, \cite{pdb}, namely the 
effective mixing angle, $\overline{sin}^2_f \hat{\theta} (M_Z) {\scriptstyle (\overline{MS})} \, = \, 0.23152(14)$, the value of the strong coupling constant: $\alpha_s (M_Z) \, = \, 0.1176(20)$ and the electromagnetic coupling $\alpha_{em}\,$.
Note that the experimentally measured effective mixing angle $ \overline{sin}^2_f \hat{\theta} $ is not the same with $ \sin^2 \theta_w $ appearing in (\ref{allit}), which is defined as ratio of couplings in the $\overline{DR}$ scheme. Actually, the two are related by 
$$
\overline{sin}^2_f \hat{\theta} \,=\,\sin^2 \theta_w \,( 1 + \Delta k_f \,) \, ,
$$
where $ \Delta k_f \,$ is calculated by the effective $Z f \overline{f} \,$ coupling. 

The minimization conditions are solved with all one-loop effective potential corrections and the dominant two-loop QCD and top Yukawa corrections taken into account. The value of the $| \mu |$ parameter is then determined by the minimization conditions and $\tan \beta \,$ is input. Therefore, apart from the particular treatment of HETs, the procedure is the standard one encountered in the constrained MSSM models. 
As far as proton decay is concerned, 
for given SUSY inputs, $m_0, \, M_{1/2},\, A_0, \, \tan \beta$, we derive $M_{eff}$, from (\ref{yan_ours}), and  $B_i$'s, in eq.(\ref{bi_prtn}), which both affect the proton decay. Then proton lifetime is calculated and it provides an additional constraint. 
Since the dependence of $M_{eff}$ on HETs is explicit only on the quantity $ x \,$,  defined by eq. (\ref{sh_param}), to facilitate the analyses we shall pick slices of fixed values of $ x \,$, in the space of random points, within which $M_{eff}$ is almost constant.

\section{Numerical Analysis - Results} \label{results1}

We follow the procedure described in Sec.\ref{secproton} and \ref{sechet}, in order to delineate the acceptable parameter region of the model at hand, which complies  with 
electroweak precision data and proton decay constraint. Satisfaction of the experimental  bounds on the strong coupling constant  \astrong, by itself,  imposes severe constraints, as we shall see, in conjunction with precision measurements and unification conditions. 

In order to facilitate the analysis, we generate random samples of randomly generated  points $\vec{c}$ 
for which $\, M_{GUT}\,$ is fixed. In our analysis, we present results for $\, M_{GUT} = 2 \,\cdot \, 10^{16} \, GeV$  but higher values are not excluded, yielding qualitatively similar results. However, perturbativity limits on Yukawa and gauge couplings poses upper bounds on higher $\, M_{GUT}$ values and hence such large values are not considered. Besides, we select points defining \textquotedblleft slices\textquotedblright  in the space of $\vec{c}$ vectors for which the ratio $x \,$ defined in 
eq.   (\ref{sh_param}) is $x \,=5\,$. In fact, we have found that a larger $x\,$ satisfies the proton decay constraint easier, while a smaller one fails on both proton decay and \astrong constraints.  
In the following, for a more clearer presentations, only 1000 points are displayed in each Figure.

In general, the effective weak - mixing angle, denoted in the Figures by $s_f$, takes values with error less than $3\sigma$ over all the parameter space, but the strong coupling constant 
prefers rather low values of SUSY breaking parameters. This is expected since for high values  SUSY is absent, due to decoupling, and, in this case, we deal with a conventional GUT model for which gauge coupling constant unification is hard to achieve.  On the other hand, high  SUSY breaking parameters and, in particular $M_{1/2}\, $, which affects wino masses and in turn Eq. (\ref{bjila}), shrink the range  of the allowed values of $M_{eff}$, by proton decay bounds, leaving out a small number of the initial randomly generated points that pass successfully the proton decay and precision data tests. 
This is the reason only relatively small values for the SUSY breaking parameters are considered, 
$\, m_0, \, M_{1/2} \, \leq \, 1.5 \, TeV   \,$.

\subsection{The \astrong  and proton decay constraints}
\label{astrongproton}

\begin{figure}[h!]  
\vspace{-1.5cm}
\begin{center}
\rotatebox{0}{\includegraphics[height=11cm,width=14cm]{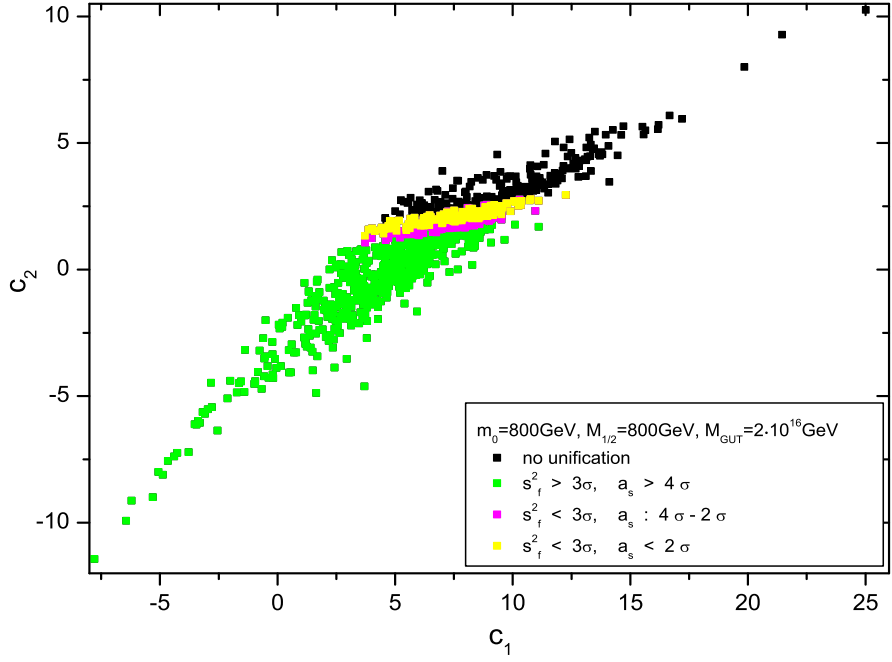}} 
\vspace{-1cm}
\caption[]{\small  In this Figure $\tan \beta \, = \, 10$ and $ A_0 \, = \, 100$ GeV, with $m_0, \, M_{1/2}$ and $ \mgut$ as displayed.  
For a detailed description of the Figure see main text. }
\label{astr80080016c2}  
 \end{center}
\end{figure}

\begin{figure}[h!] 
\vspace{-2cm}  
\begin{center}
\rotatebox{0}{\includegraphics[height=11cm,width=14cm]{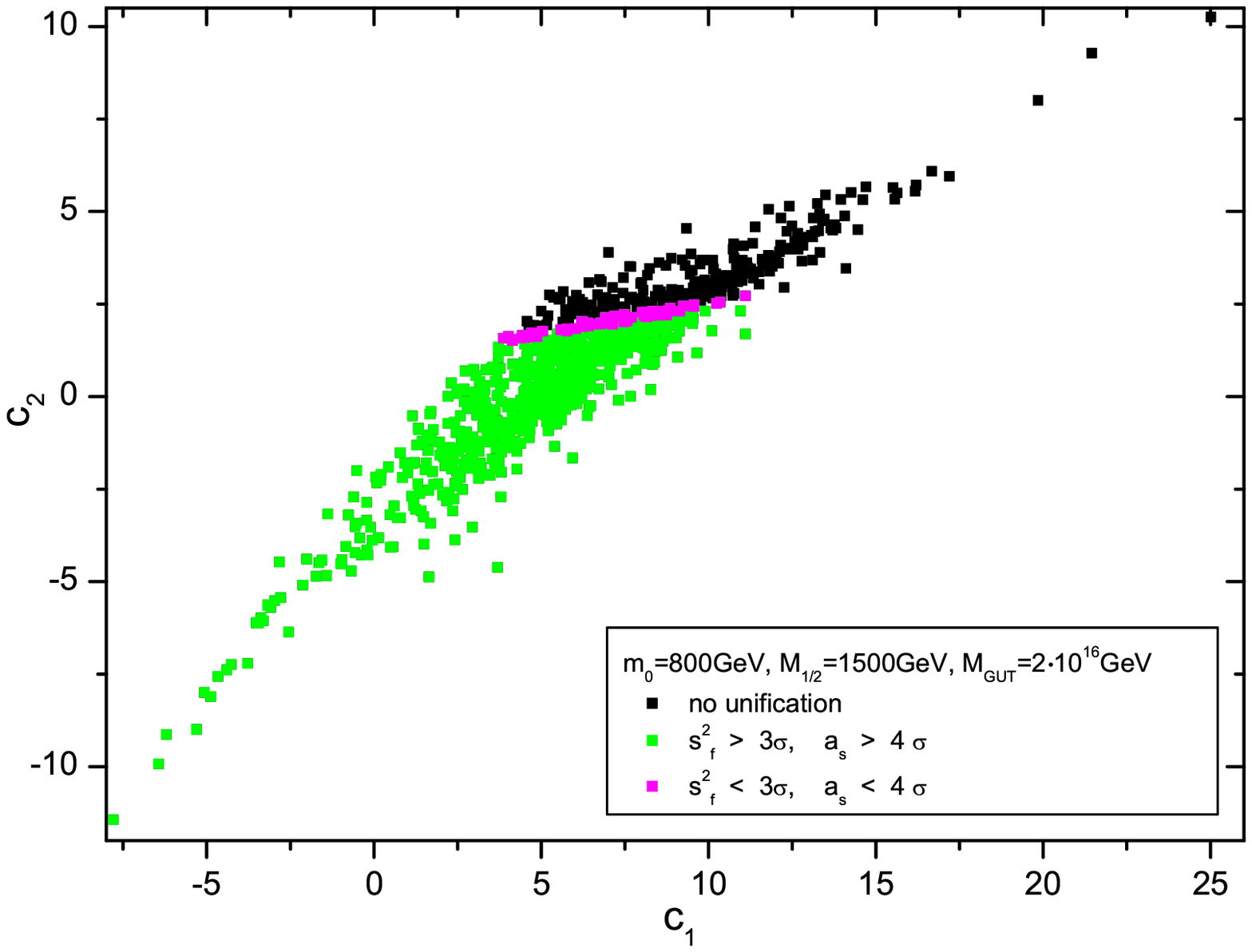}} 
\vspace{-1cm}
\caption[]{\small   As in Figure \ref{astr80080016c2}.}
\label{astr800150016c2}  
 \end{center}
\end{figure}

\begin{figure}[]  
\vspace{-2cm} 
\begin{center}
\rotatebox{0}{\includegraphics[height=11cm,width=14cm]{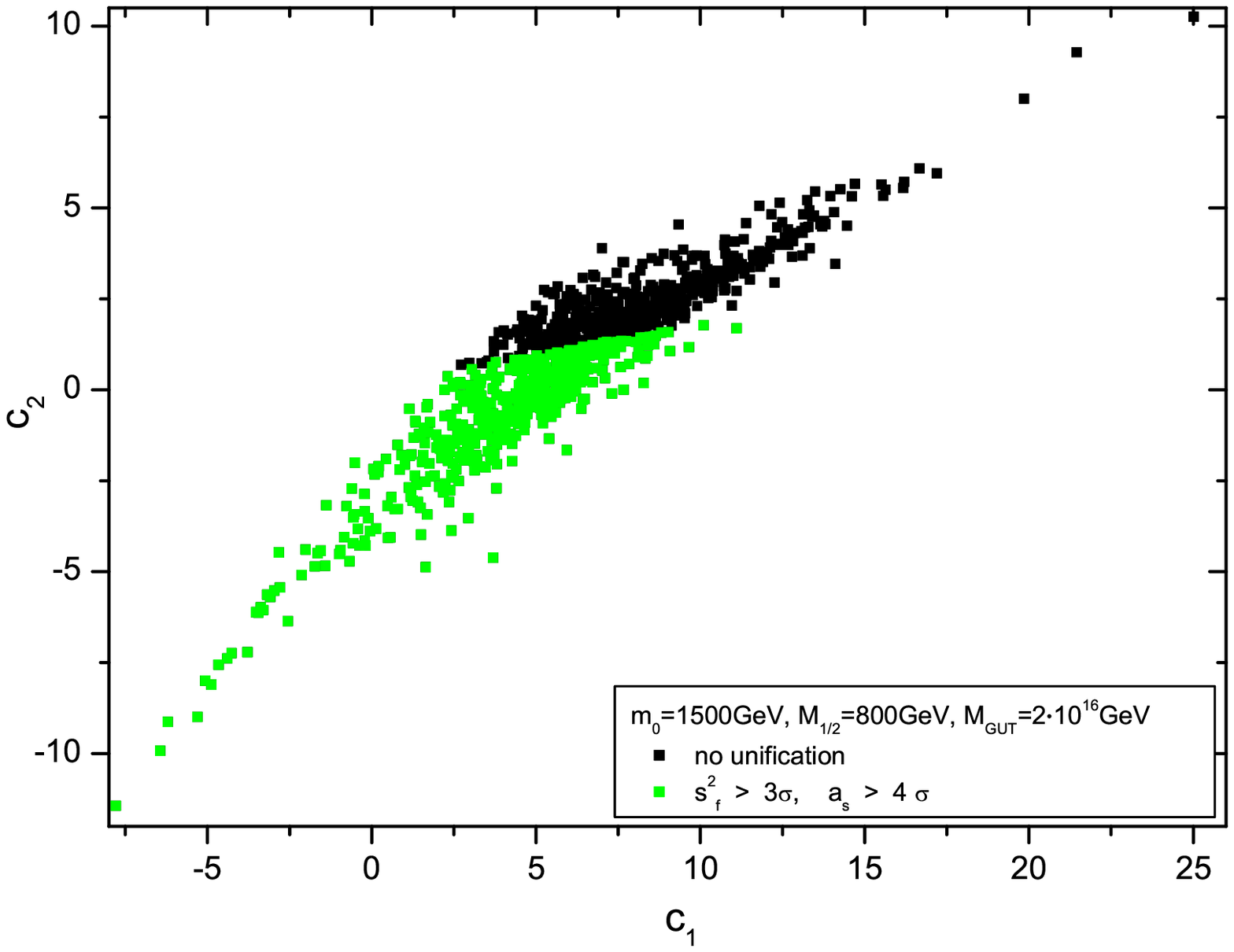}} 
\vspace{-1cm}
\caption[]{\small   As in Figure \ref{astr80080016c2}.}
\label{astr150080016c2}  
 \end{center}
\end{figure}

\begin{figure}[] 
\vspace{-2cm}  
\begin{center}
\rotatebox{0}{\includegraphics[height=11cm,width=14cm]{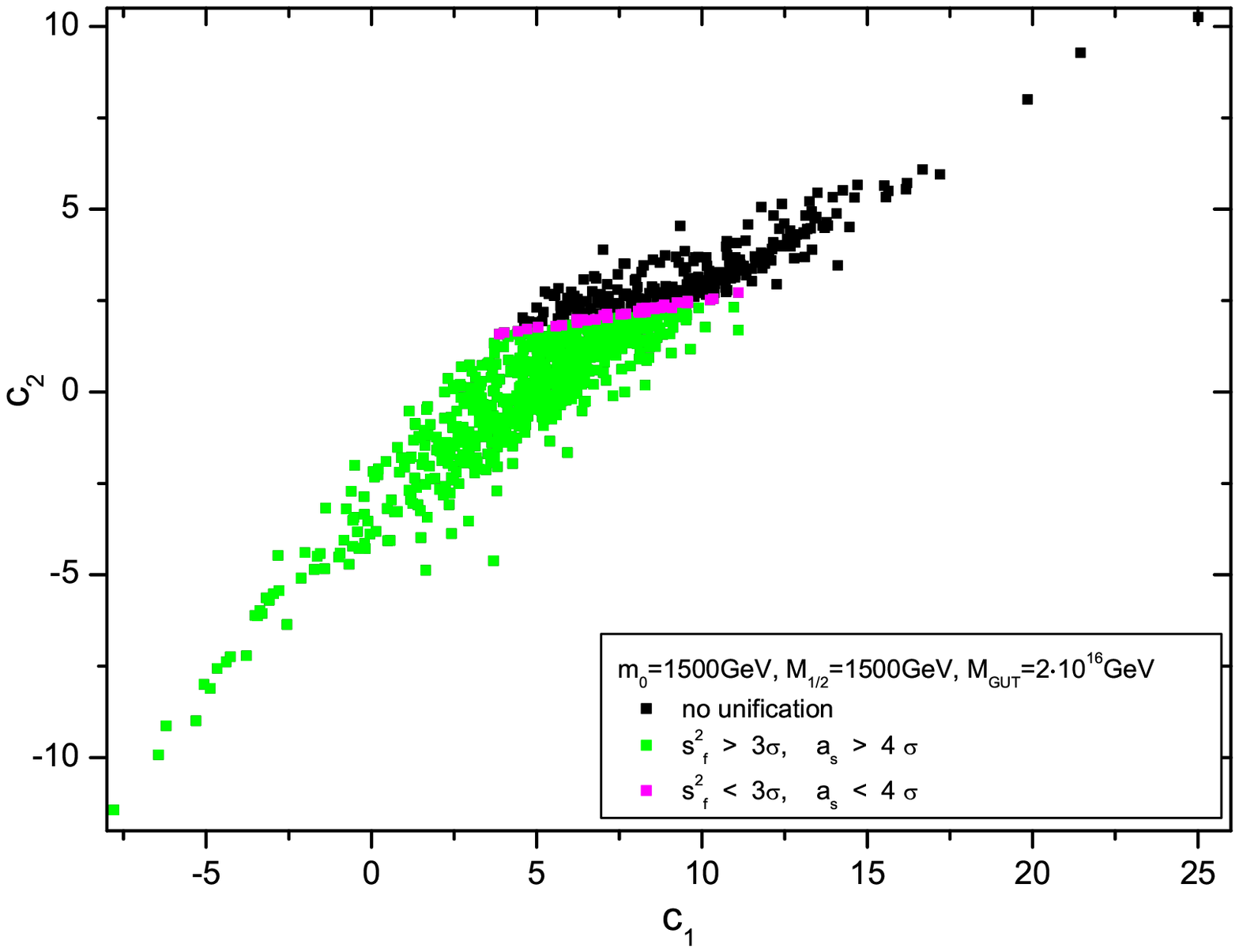}} 
\vspace{-1cm}
\caption[]{\small   As in Figure \ref{astr80080016c2}.}
\label{astr1500150016c2}  
 \end{center}
\end{figure}

We come to the point of examining the dependence of the  strong coupling constant \astrong  on the supersymmetric and HET parameters. We first observe its variation with changing $m_0$ and $ M_{1/2}$. In Figure \ref{astr80080016c2}-\ref{astr1500150016c2}, we display the pairs of $c_1, c_2$ as they are randomly generated.
The value of the unification scale has been taken $\, M_{GUT} = 2 \,\cdot \, 10^{16} \, GeV$ and on naturalness reasons all high energy parameters having dimension of mass are randomly generated with values differing from $\, M_{GUT}$ by at most three orders of magnitude. Although these parameters have been randomly generated from the independent parameters of the model, in the way prescribed earlier,  they are correlated as shown clearly in the Figures. Therefore, successful points ought to be within the diagonal stripe displayed in these Figures.
 
The black points represent those pairs of $c_1, c_2$ that fail even to give unification at the quoted \mgut, after the 2 - loop running of the  RGEs. For the green points, unification has been achieved but the value of \astrong is more than 
$4\sigma$ away. The magenta points yield \astrong with error smaller than $4\sigma$, and the yellow region is the subset 
of magenta points corresponding to values of \astrong with the smallest possible error $< 2\sigma$. 

\begin{figure}[h!]  
\vspace{-2cm}
\begin{center}
\rotatebox{0}{\includegraphics[height=11cm,width=14cm]{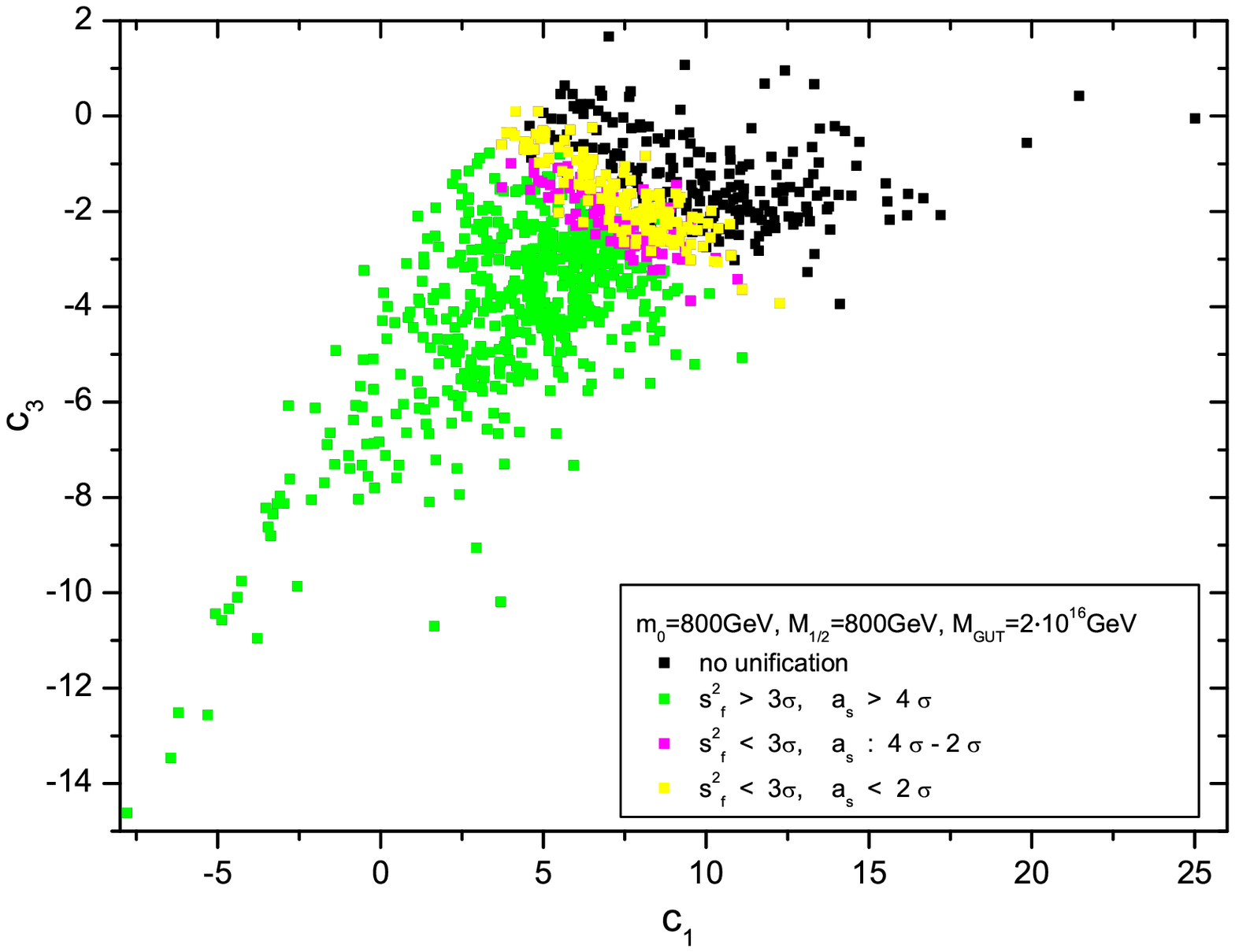}} 
\vspace{-1cm}
\caption[]{\small  As in Figure \ref{astr80080016c2}.}
\label{astr80080016c3}  
 \end{center}
\end{figure}

In Figures \ref{astr80080016c2}  to  \ref{astr1500150016c2}, we vary the input values of $m_0$ and $ M_{1/2}$, as shown on each Figure, by increasing either one, or both, of $m_0\,$, $ M_{1/2}$, but keeping $\mgut$, $\tan \beta$ and $ A_0 $ fixed. In Figure \ref{astr80080016c3}, we display the analog of Figure \ref{astr80080016c2} but with $c_2$ replaced by $c_3$. 
A correlation of $c_1$ and  $c_3$ is also observed with points that are more widely scattered. 
%
%

By raising the values of $m_0$ and $ M_{1/2}$, we have observed a considerable decrease of the number of random sample points that result in an acceptable \astrong value. The yellow region shrinks and shifts slightly towards smaller values of $c_2$ and eventually disappears, as is the case in the displayed Figures \ref{astr800150016c2}  to  \ref{astr1500150016c2}. In those cases, the deviation of the theoretical from the experimental value of \astrong never becomes less than $ 2\sigma\,$. In Figure \ref{astr150080016c2}, by increasing the value of $m_0$, keeping $ M_{1/2}$ = $800$ GeV as in Figure \ref{astr80080016c2}, the yellow and the magenta regions disappear, which means that the predicted theoretical values for \astrong are outside the experimental limits. 
Therefore, to be within experimental limits, the values of $ M_{1/2}$ and primarily $m_0$ should be kept central to small. 
The \astrong constraint is best met in a region of the $m_0$ - $ M_{1/2}$ plane bounded by values of $m_0$ up to $1000$ GeV and $ M_{1/2}$ up to $1300$ GeV. For larger values of $m_0$ up to $1400$ GeV, we should choose $ M_{1/2}$ between $900$ and $1300$ GeV at most and for larger values of $m_0$ up to $1500$ GeV, we should go for small $m_0$ around $500$ GeV.
This conclusion was rather expected since for large values of $m_0$ and $ M_{1/2}$, exceeding $1$ TeV, the MSSM and the Standard Model, due to decoupling,  
give almost the same predictions for the unification of coupling constants, where SM fails to yield a satisfactory unified picture.

Based on the most recent experimental bound on proton lifetime (\ref{plifetime}), the lower bound on its lifetime translates into a lower bound on the mass parameter \meff that controls the proton decay width. This, denoted by \meffexp,  can be extracted from eq. (\ref{width}) and it is defined below. We perform these calculation for every randomly generated pair of  $c_1, c_2$ and the constraint of proton decay is satisfied provided that 
\beq
\meffth > \meffexp \, \equiv \, \beta_p \, | A | \, \sqrt{  \tau_b \,  C \, \sum_i \, {| B_i | }^2      },
\label{prtconstr}
\eeq
where $\tau_b$ is the bound in (\ref{plifetime}) and \meffth is read from (\ref{yan_ours}). 

\begin{figure}[h!]  
\vspace{-1.5cm}
\begin{center}
\rotatebox{0}{\includegraphics[height=11cm,width=14cm]{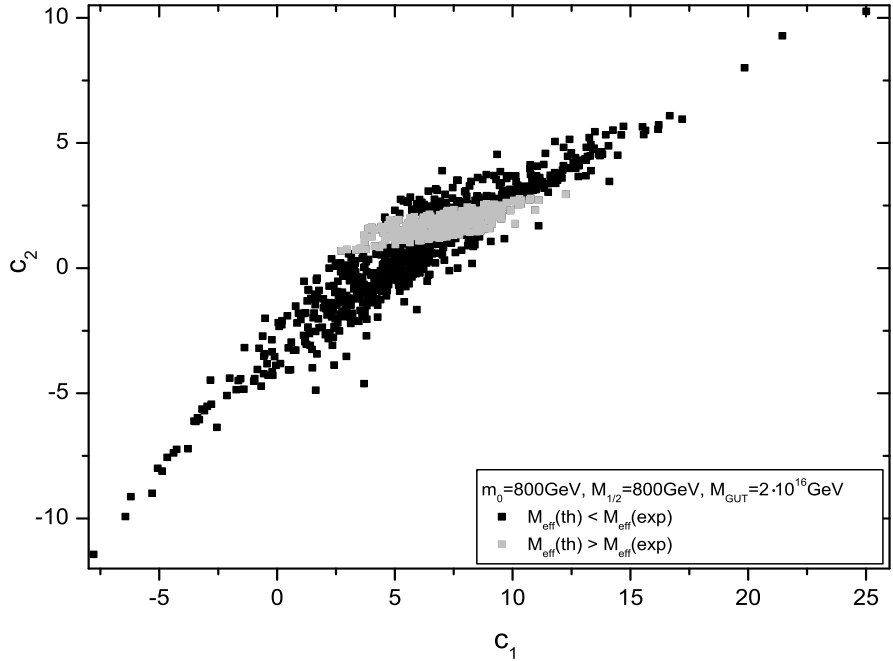}} 
\vspace{-1cm}
\caption[]{\small  In this Figure $\tan \beta \, = \, 10$ and $ A_0 \, = \, 100$ GeV, with $m_0, M_{1/2}, \mgut$ as quoted. For a detailed description of the Figure see main text.}
\label{prt80080016c2}  
 \end{center}
\end{figure}

Running the RGEs we find points for which (\ref{prtconstr}) is satisfied and the resulting \meffth yields a proton lifetime in the range $10^{34} - 10^{37}$ years.
In Figure \ref{prt80080016c2}, we illustrate this constraint for the case of $m_0 \, = \,M_{1/2} = \, 800$ GeV, by coloring in gray the pairs of $c_1, c_2$ which fulfill (\ref{prtconstr}) and in black the rest of those. 
These points overlap with the majority of the points that yield gauge coupling unification with values of \astrong within the 
 $ 2\sigma\,$ experimental range.
Raising of $m_0$ and/or $ M_{1/2}$ results to an homogeneous reduction of the gray region around a central point. Besides, the unilateral increase of $m_0$ keeping $ M_{1/2}$ rather small leads to experimentally unacceptable proton decay rates.
%
%

The results depend on the unification scale \mgut, which is an input in our analysis. For \astrong, raising \mgut by almost one order of magnitude ( $\mgut \, = \, 9 \cdot 10^{16} \mathrm{GeV}$ ) broadens the yellow region of Figure \ref{astr80080016c2} which now includes smaller values for $c_1, c_2$. We note that almost half of the yellow points yield \astrong within $1\sigma$ error. Taking the same high value for \mgut, in the case of the proton decay constraint, results to the same behavior. In fact, the gray region of Figure
\ref{prt80080016c2} moves towards smaller values of $c_2$, while it reaches a wider range of values for $c_1$. It is interesting to note that half of the points which bring unification of the coupling constants fulfill the proton decay constraint for $\mgut \, = \, 9 \cdot \, 10^{16} \mathrm{GeV}$, while for $\mgut \, = \, 2 \cdot \,10^{16} \mathrm{GeV}$ the corresponding rate is 40\%. Thus, by pushing \mgut to higher value provides easier satisfaction of the constraints. As far as $ A_0 $ is concerned, this parameter does not seem to play a significant role in our analysis.

Our findings depend also on the value of $\tan \beta$. A change of $\tan \beta$ from 10 to 45 causes a small decrease, ranging from 5 to 15\%, in the number of points which succeed to give unification and, at the same time, shrinks their range of output values. A likely explanation for this is that a high value of $\tan \beta$ increases the chance of a Landau pole to appear during the running of the RGEs. Also, the raise of $\tan \beta$ bounds the points which give \astrong with error more than $4\sigma$ to have only positive $c_1$ and only negative $c_3$. On the other hand, with a large $\tan \beta$ the number of points which give \astrong with error less than $4\sigma$ (or $2\sigma$ for $m_0 \, = \, M_{1/2}$ = $800$ GeV) slightly increases.
As far as the proton decay constraint is concerned, the points which satisfy (\ref{prtconstr}) show a considerable decrease, which starts from 22\% for $m_0 \, = \, M_{1/2} \, = \, 800$ GeV and reaches a 100\% for $m_0 \, = \, 1500$ GeV and $M_{1/2} \, = \, 800$ GeV. This was rather expected since $B_i$ in (\ref{prtconstr}) depends on $\frac{1}{\sin 2\beta}$. Hence, a change of $\tan \beta$ from 10 to 45 quintuples or so the values of \meffexp leaving, at the same time, the values of \meffth almost unchanged. Therefore, the chance of satisfying (\ref{prtconstr}) decreases. Overall, a change of $\tan \beta$ from 10 to 45 makes the satisfaction a harder task.

\subsection{Convergence of results}

In section \ref{sechet}, we pointed out that, in general, there are differences between the final points
$\vec{c}_{fin} \, = \, \{c_{1}^{fin}, c_{2}^{fin}, c_{3}^{fin}, M_{L}^{fin} \}$, stemming from the running of the RGEs (output values), and the initial (input values)  $\vec{c}_{in} \,=\, \{ c_1^{in}, c_2^{in}, c_3^{in}, M_L^{in} \}$, generated from the random sample, owning to the corrections imposed to achieve unification at the given \mgut. For the case of $c_1$ and $c_2$, there is a close agreement between the input and output values but the same cannot be asserted for $c_3$ as discussed in section \ref{sechet}. Actually, this deviation is expected since the $c_3$ parameter is strongly correlated to \astrong and is forced to make the strong coupling be compatible with the unification scale that is determined by the couplings $\alpha_{1,2} \,$. This is implemented by the shift of eq.  (\ref{c3}), resulting to $c_3^{fin}$, which is always towards higher values comparing to input $c_3^{in} \,$. This is also the case for $c_{1,2}^{in} \,$ and $M_L^{in}\,$, which are also shifted towards higher values but to a much lesser extend.

Focusing only on the ability of the random sample points to accomplish unification of the gauge couplings, without any other experimental constraint, an overlapping region of both input and output points must exist for the model to be successful. We have found that the gap between the initial and final points augments by increasing $m_0$ and by reducing \mgut.

\begin{figure}[h]  
\vspace{-2cm}
\begin{center}
\rotatebox{0}{\includegraphics[height=13cm,width=16cm]{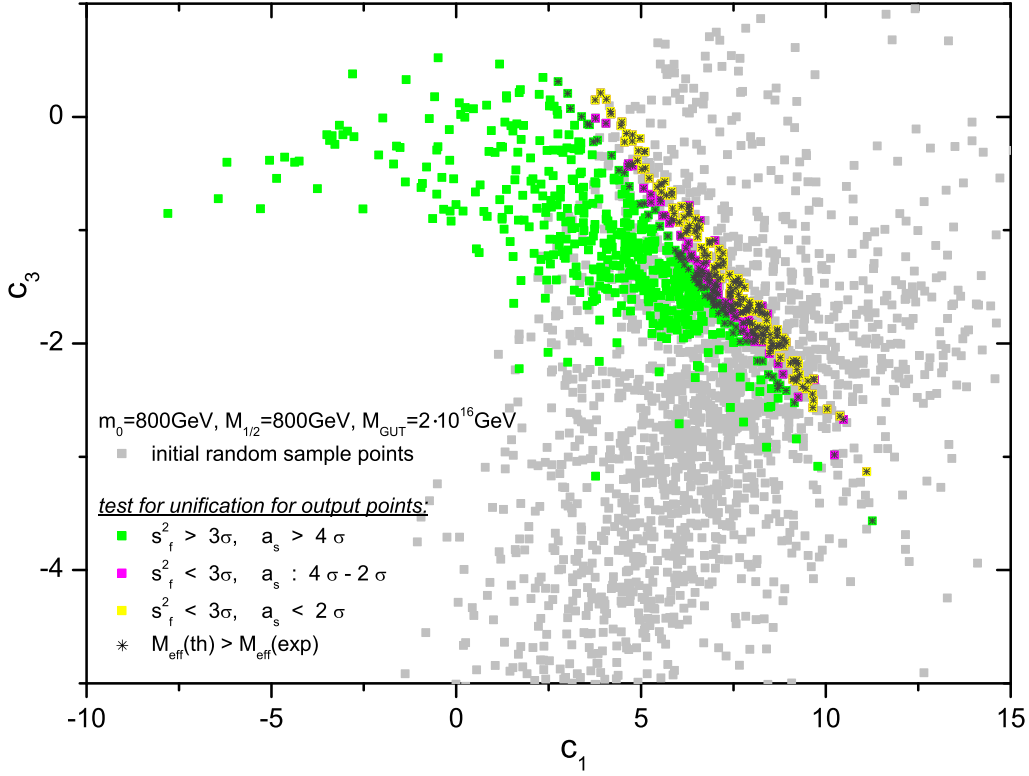}} 
\vspace{-1.5cm} 
\caption[]{\small  $\tan \beta \, = \, 10$ and $ A_0 \, = \, 100$ GeV, with $m_0, M_{1/2}, \mgut$ as quoted. For a detailed description of the Figure see main text.}
\label{outputs_astr_c1c3_zoom}  
 \end{center}
\end{figure}

The situation becomes more constrained if we simultaneously demand satisfaction of both \astrong and proton decay constraints. 
For instance in Figure \ref{outputs_astr_c1c3_zoom} we display in gray the points of $c_1, c_3$ as they are randomly generated. Actually, this represents a magnification of the central and upper area of Figure \ref{astr80080016c3}. The green points represent the output values $c_{1}^{fin}, c_{3}^{fin}$ of the gray input points $c_1^{in}, c_3^{in}$ which yield  values of \astrong with error higher than $4\sigma $. The magenta points are the corresponding points which yield \astrong with smaller error, between $4\sigma $ and $2\sigma $, while the yellow points give \astrong with error less than $2\sigma $. The gray crosses designate those output points that satisfy the proton decay constraint. 

We perform a "second" run using as inputs the elements of the previously produced output groups. 
In this run, we first check whether these points are mapped to themselves as they should.  These are likely to be successful points provided they are subset of the initial randomly generated points. Obviously, only a subset of those survives, if it does at all, and this comprises the set of successful points.  In order to check numerically whether a final point belongs to the set of the initial points, we define the "distance" $ \chi_i \, $ of the point $ \vec{c}_{fin} \,$ from any $ \vec{c}_{in}^{\,i} \,$ of the randomly generated points
$$
\chi_i \equiv  \, \left|\frac{{c}^{fin}_1 - {c}^{\, in, \, i}_1}{{c}^{\,in, \, i}_1} \right| \,+\, 
\left|\frac{{c}^{fin}_2 - {c}^{\, in, \, i}_2}{{c}^{\,in, \, i}_2} \right| \,+\,...\,
\left|\frac{{c}^{fin}_5 - {c}^{\, in, \, i}_5}{{c}^{\,in, \, i}_5} \right| \, .
$$
The minimum of these $\,\chi = min \{ \chi_i's \} \, $ defines the distance of $\, \vec{c}_{fin} \,$ from the set of the randomly generated points and if zero it means that the final point coincides with one of the random points that were initially created. In practice, the smallness of $\chi$ indicates that the particular point is successful in the sense that it belongs to  the \so model and in addition it agrees with the experimental  results.

In our analysis and with one million random points, we have found that points with  $\chi \leq  0.1$ are acceptable by the model. 

This analysis certainly depends on the SUSY inputs.  In the $m_0$ - $ M_{1/2}$ plane, successful points are found  for values of $m_0$ and $ M_{1/2}$ reaching roughly $1000$ GeV. If one keeps the higher end of $ M_{1/2}$ values constant, the restrictions are satisfied altogether for values of $m_0\,$ up to approximately $1200$ GeV. If we loosen the restriction for $\chi$ to $\chi \leq  0.2$, keeping the other two constraints untouched, our successful region matches the one described in subsection \ref{astrongproton}. This findings supports not only our method but also the \so model we have followed on the ground of satisfying \astrong and proton decay constraints.

\section{Conclusions} \label{conclusions}

We have presented a new approach towards treating in a collective way the large number of HET encountered in GUTs quantities by using a random sample technique. According to this technique the large number of the GUTs parameters, defining the high energy sector and hence the threshold masses, are mapped to a few properly defined parameters encoding all the information associated with the renormalization group running of the various quantities involved from the unification scale down to electroweak energies. 
This method is simple and efficient since : 
\begin{enumerate}
\item 
It avoids unnecessary runnings, occurring in the conventional scheme, where for each point in the multidimensional parameter space one has to solve numerically the RGEs. 
\item
The parameter space is mapped to properly defined quantities, associated with the running of the gauge couplings, which collectively include enough information through which regions favoured by precision data on gauge couplings are easily located. 
\item
The RG equations are solved without the inclusion of High Energy Thresholds and their effect is duly taken into account by changing appropriately the boundary conditions at the lowest of the high energy thresholds and at the Unification scale. 
\item
This scheme is fast and accurate if the lowest, $M_L$,  and the highest, $ M_{GUT}$, of the high energy threshold are separated by at most three orders of magnitude, $ \,M_L / M_{GUT} \,< \,10^{-3}  \, $.
\end{enumerate}

We applied this technique to an \so-based model where only five parameters ($c_{1,2,3} \, \tan \theta$ and $M_L$) embody all information associated with the HETs. Our analysis shows that this method is both convenient and efficient. We test it and demonstrate that 
there exists a confined region of $c_i$s values which yields results in agreement with the limitations imposed by precision data and coupling constant unification, as well as proton decay constraints. 
These regions are favored by central to small values of $m_0$ and $ M_{1/2}$ in the  the region of $500 \,$ GeV $- 1.5 \,$ TeV and are suppressed for large $\tan \beta$ due mainly to proton decay constraints.  The effect of the common trilinear coupling $ A_0 $ is small provided it lies in the TeV range.

\section*{Acknowledgements}
The author is grateful to A.B. Lahanas for extensive discussions, critically reading 
the manuscript and continuous support during this effort.
The author wishes to acknowledge support  by the European Union through the fp6  Marie-Curie Research and Training Network Heptools (MRTN-CT-2006-035505). She  acknowledges also partial support from  the University of Athens Special Research Account.



\end{document}